\documentclass[lettersize,journal]{IEEEtran}
\usepackage{amsmath,amsfonts}
\usepackage{algorithmic}
\usepackage{algorithm}
\usepackage{array}
\usepackage[caption=false,font=normalsize,labelfont=sf,textfont=sf]{subfig}
\usepackage{textcomp}
\usepackage{stfloats}
\usepackage{url}
\usepackage{verbatim}
\usepackage{graphicx}
\usepackage{cite}
\usepackage{color}

\begin{document}

\title{Decoding and Engineering the Phytobiome Communication for Smart Agriculture}

\author{
    Fatih Gulec, Hamdan Awan, Nigel Wallbridge, Andrew W. Eckford
    \thanks{Fatih Gulec is with the University of Essex, UK. Hamdan Awan is with Munster Technological University, Ireland. Nigel Wallbridge is with Vivent SA, Switzerland. Andrew W. Eckford is with York University, Canada.}%
    %
  }



\maketitle

\begin{abstract}
Smart agriculture applications, integrating technologies like the Internet of Things and machine learning/artificial intelligence (ML/AI) into agriculture, hold promise to address modern challenges of rising food demand, environmental pollution, and water scarcity. Alongside the concept of the phytobiome, which defines the area including the plant, its environment, and associated organisms, and the recent emergence of molecular communication (MC), there exists an important opportunity to advance agricultural science and practice using communication theory. In this article, we motivate to use the communication engineering perspective for developing a holistic understanding of the phytobiome communication and bridge the gap between the phytobiome communication and smart agriculture. Firstly, an overview of phytobiome communication via molecular and electrophysiological signals is presented and a multi-scale framework modeling the phytobiome as a communication network is conceptualized. Then, how this framework is used to model electrophysiological signals is demonstrated with plant experiments. Furthermore, possible smart agriculture applications, such as smart irrigation and targeted delivery of agrochemicals, through engineering the phytobiome communication are proposed. These applications merge ML/AI methods with the Internet of Bio-Nano-Things enabled by MC and pave the way towards more efficient, sustainable, and eco-friendly agricultural production. Finally, the implementation challenges, open research issues, and industrial outlook for these applications are discussed.

\end{abstract}
\vspace{-0.5cm}

\section{Introduction}
\IEEEPARstart{T}{he} world population is estimated to reach nearly $10$ billion by $2050$, and accordingly, the demand to increase agricultural production is rising. While the $70\%$ of the world's water is consumed in agriculture, water scarcity is increasing every year due to climate change. In addition, agrochemicals such as pesticides and fertilizers are used to improve crop production efficiency. The excessive use of these agrochemicals is harmful to the environment and living things from humans to bees. Furthermore, their delivery efficiency is low, e.g., less than $0.1\%$ of applied pesticides reach their target pests in the USA \cite{vega2020nanoscale}. These issues raise concerns about public health, biodiversity loss, and food security. Hence, sustainable, efficient, and eco-friendly methods are needed in agriculture.


To improve agricultural production efficiency, smart agriculture applications have emerged thanks to various technologies such as the Internet of Things (IoT),  robotics, and machine learning/artificial intelligence (ML/AI) \cite{shaikh2022recent}. Via these applications, plants and their environment are monitored, typically focusing on environmental parameters. In addition, smart agriculture applications include plant monitoring with cameras and optical sensors to detect plant stress such as pest attacks, infections, and nutrient deficiencies. These collected sensor data can be transferred to monitoring centers using available communication technologies such as WiFi, ZigBee, or cellular networks, and processed to support decisions on resource management, including irrigation and agrochemicals.

The drawback of these applications is their delayed detection of plant stress. For instance, a bacterial infection is typically detected only after visual symptoms appear, which increases crop loss. As a solution, electrophysiological signals, which are used by plants to internally communicate information, emerged for early stress detection \cite{najdenovska2021identifying}. However, this type of monitoring treats the plant as a black box, lacking insight into internal mechanisms. Moreover, current smart agriculture applications do not consider the plant and its environment as an ecosystem but only measure and extract information from individual components. 

Nevertheless, a plant is not independent of its environment and organisms around it. To establish a holistic perspective, the phytobiome, which is defined as the area including the plant, its environment, and organisms within and around the plant, has been proposed \cite{trivedi2020plant}. However, the intricate communication among the organisms in the phytobiome is far from being fully understood in terms of modeling and utilizing it for agricultural applications. This communication is essential for understanding the relationships among the organisms and for identifying ways to exploit them effectively. Recent progress in biophysics, molecular communication (MC), and mathematical biology highlight that information gathering and processing are fundamental to an organism's fitness, i.e., its ability to adapt and survive \cite{donaldson2010fitness}. Hence, there is a need to model and design the phytobiome communication in a unified framework.

\begin{figure*}[t]
	\centering
	\includegraphics[width=0.75\textwidth]{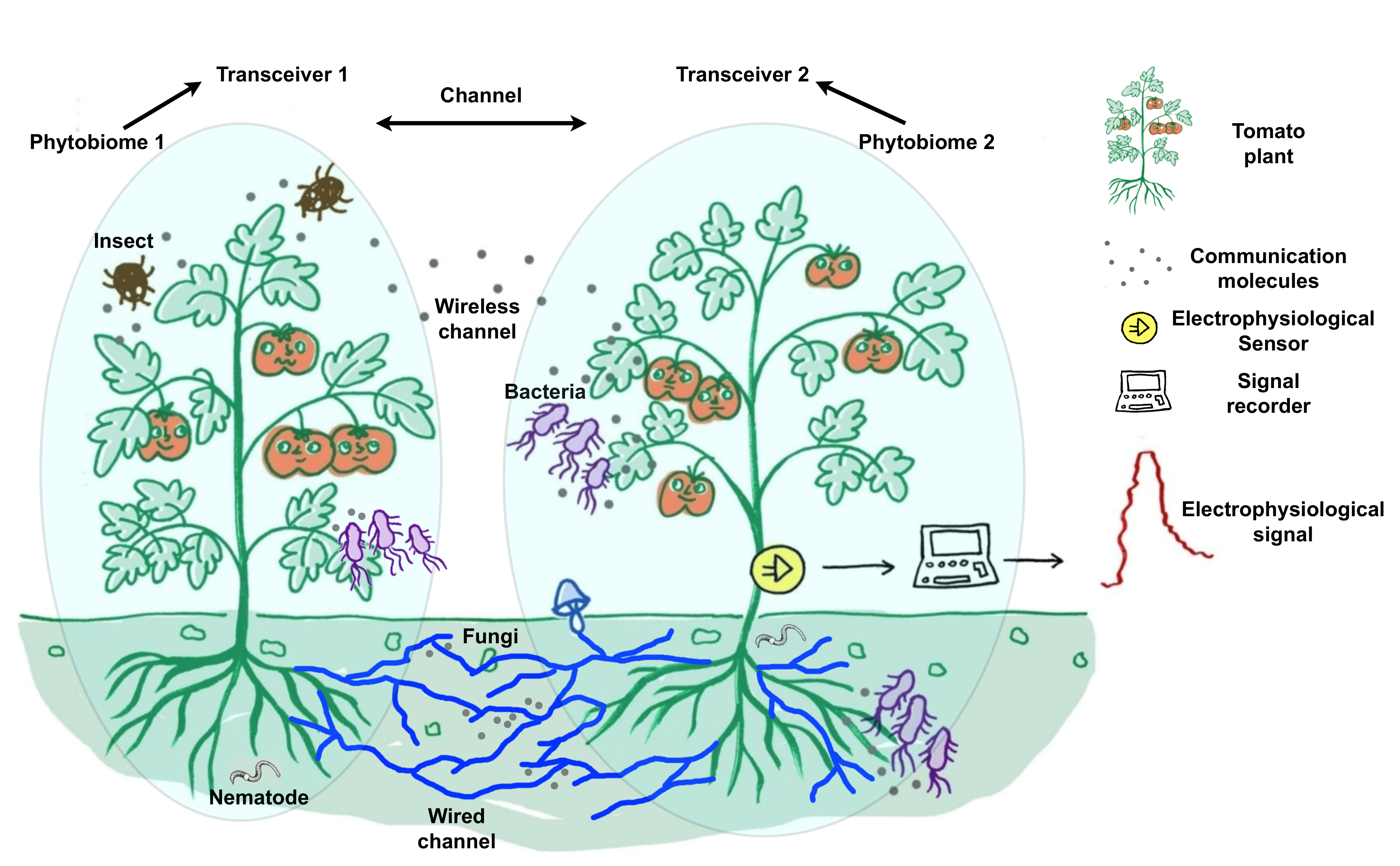}
	\caption{Phytobiome and its components.}
	\label{Phytobiome}
 \vspace{-0.3cm}
\end{figure*}

As illustrated in Fig. \ref{Phytobiome}, phytobiome members such as bacteria, fungi, nematodes, and insects interact with plants by exchanging molecules, i.e., MC.
MC emerged as a promising field that blends communication engineering with biology, biophysics, and chemistry. It was proposed as a paradigm for the communication among nanomachines (NMs) to accomplish advanced tasks such as targeted drug delivery \cite{farsad2016comprehensive}. Moreover, the MC concept gave rise to the Internet of Bio-Nano Things (IoBNT), where nanonetworks consisting of NMs are remotely controlled through the Internet. Importantly, MC has also been utilized to understand natural phenomena such as infectious disease spread  \cite{gulec2022mobile} and information propagation in plants \cite{awan2019communication}.  

In this article, we motivate the use of the communication engineering perspective to decode and engineer phytobiome communication, and employ this approach for smart agriculture applications. Unlike conventional IoT-based systems monitoring environmental parameters around plants, our approach focuses on the plant as an active information source and its communication with organisms around it. Our work leverages this viewpoint to propose a multi-scale communication framework using principles of MC theory spanning from intercellular communication in microscale to the communication among phytobiomes in macroscale. While intra- and inter-plant communication as well as cross-organism communication have been studied separately in plant biology, to our knowledge, this is the first attempt to unify them under a multi-scale communication engineering framework, linked to measurable electrophysiological signals. This also leads to new application concepts using ML/AI and IoBNT in agriculture. To this end, we propose to consider the phytobiome as a communication network. First, we review the current understanding of phytobiome communications and present a multi-scale communication framework. Second, we review our previously proposed approach that uses MC to model electrophysiological signals in plants, including its experimental validation. Third, novel smart agriculture applications enabled by engineering phytobiome communication are proposed. These applications include phytobiome monitoring, building on prior work using ML/AI for stress detection from plant electrophysiology, targeted agrochemical/gene delivery, smart irrigation, and engineering intra- and inter-phytobiome communication. The first three applications are merged for stress diagnosis, optimized water use, and targeted delivery of agrochemicals, where the plant autonomously manages itself. To this end, we propose integrating ML/AI methods into the use of IoBNT. To guide implementation, we outline associated challenges and open research issues, and conclude with an industrial outlook.


\section{Decoding the Phytobiome Communication} \label{DPC}
A phytobiome is a network wherein a plant has intricate connections with various organisms as illustrated in Fig. \ref{Phytobiome}. This network includes viruses and organisms across various kingdoms, encompassing bacteria, archaea, protists, fungi, and animals \cite{leach2017communication}. Communication through molecular and electrical signals is imperative for the survival and reproductive success of phytobiome inhabitants. In this section, we first review the state-of-the-art of this communication in two categories: intra-kingdom and inter-kingdom. Subsequently, we explain how the communication engineering perspective can be applied to model phytobiome communication in various scales, including experimental validation of our MC model for electrophysiological signals.

\subsection{Overview of the Phytobiome Communication}
\subsubsection{Intra-kingdom Communication} \label{IKC}
\paragraph{Plants} 
Plants are the central organisms in a phytobiome, forming relationships with other organisms.
Since they are sessile, observing their responses from a human-centric perspective is difficult. However, plants exhibit diverse responses to biotic (e.g., insect attacks), and abiotic (e.g., drought) stimuli through growth, reproduction, tropism, defense, and communication.
Furthermore, plants perceive and process stimuli to make a decision, while also considering feedback information from their responses. Among these responses, communication is vital. Within the plant, molecular and electrical signals are used to transfer information to other organs. To this end, various molecules such as lipids, ribonucleic acids (RNAs), Ca$^{2+}$ ions, reactive oxygen species (ROS) like hydrogen peroxide, and hormones such as auxin, salicylic acid (SA), and jasmonic acid (JA) are employed. For example, when a plant's leaf is attacked by a herbivore (plant-eater), other leaves are informed via JA, which triggers the secretion of toxic and repellent chemicals to prevent other leaves from being eaten. 



Plants also communicate with other plants, employing both ``wireless" and ``wired" communication as shown in Fig. \ref{Phytobiome} \cite{sharifi2021social}. The former mechanism is employed to warn nearby plants via airborne MC signals known as volatile organic compounds (VOCs). For example, a tomato plant attacked by herbivores secretes VOCs, which diffuse in the air and are received by other tomato plants to induce their defense mechanisms via JA. 
The ``wired" communication occurs via fungi around the roots. Mycorrhiza, defined as the symbiotic relationship between fungi and plants, plays an important role in enhancing the plant's capability to find water and nutrients in exchange for metabolites produced by photosynthesis. The mycorrhizal fungal networks are employed to communicate the defense mechanism via JA as well as other mechanisms such as kin recognition. Thanks to the underground ``wired" fungal networks, a forest can be called a ``wood wide web".


\paragraph{Bacteria, Fungi, and Animals}
Bacteria are known as single-cell organisms with limited capabilities. However, they can accomplish sophisticated tasks through quorum sensing (QS), which is an inter-cellular MC mechanism. In QS, bacteria exchange autoinducer molecules among each other and induce mechanisms such as biofilm formation, leading to infections on plants. Like bacteria, fungi also employ QS to regulate fungal infections and growth on plant surfaces. 

Animals in the phytobiome such as bees, spider mites, or ants use MC via pheromones for various purposes such as food localization and alarm signaling. For instance, ants can follow a single line from their nests to a food source by creating ``olfactory highways", wherein they emit, perceive, and estimate the direction of pheromones among themselves.

\subsubsection{Inter-kingdom Communication}
Inter-kingdom communication involves interactions of organisms across different kingdoms. Primarily, this communication is achieved by enhancing, mimicking, degrading, and inhibiting the intra-kingdom communication molecules, such as QS molecules, by organisms from other kingdoms.
The QS mechanism plays a significant role in the symbiotic relationship of plants and bacteria.
For example, plants can ``hack" this bacterial communication by emitting rosmarinic acid, a QS mimicker molecule that binds to autoinducer receptors of bacteria 
\cite{gulec2023stochasticd}. Thus, plants induce an immature QS response and remove bacteria and their associated biofilm. In addition, VOCs emitted by bacteria and fungi can be sensed by plants to induce defense mechanisms, while plants can modulate fungal growth and mycotoxin production through secreted lipids. 

Although microorganisms such as bacteria and fungi are generally perceived as harmful to plants, they provide advantages to plants through disease resistance, stress tolerance, and nutrient acquisition \cite{trivedi2020plant}. For instance, nitrogen-fixing bacteria convert atmospheric nitrogen into a usable form for plants. Plants can recruit microorganisms to help them resist stress, e.g., drought due to climate change. This occurs through root exudates facilitating MC between plants and microorganisms.

The defense mechanism of plants can also be induced by pheromones secreted by nematodes, causing infections. These pheromones can also be sensed by fungi to prey on nematodes. In addition, plants can affect the sexual and aggregation behaviors of insects by enhancing or inhibiting insect pheromones. Furthermore, the communication between the plant and viruses, archaea, and protists is not fully recognized, although there is evidence showing that these organisms may affect the function of other microbes and the plant defense mechanisms \cite{trivedi2020plant}. Moreover, the communication can be among more than two kingdoms, e.g., MC between the plant and the bacteria secreted by the Colorado potato beetle suppresses the plant defense via JA. Next, we elaborate on a framework to clarify this complex phytobiome communication.



\subsection{Modeling the Phytobiome as a Communication Network}
\begin{figure}[hbt]
	\centering
	\includegraphics[width=0.8\columnwidth]{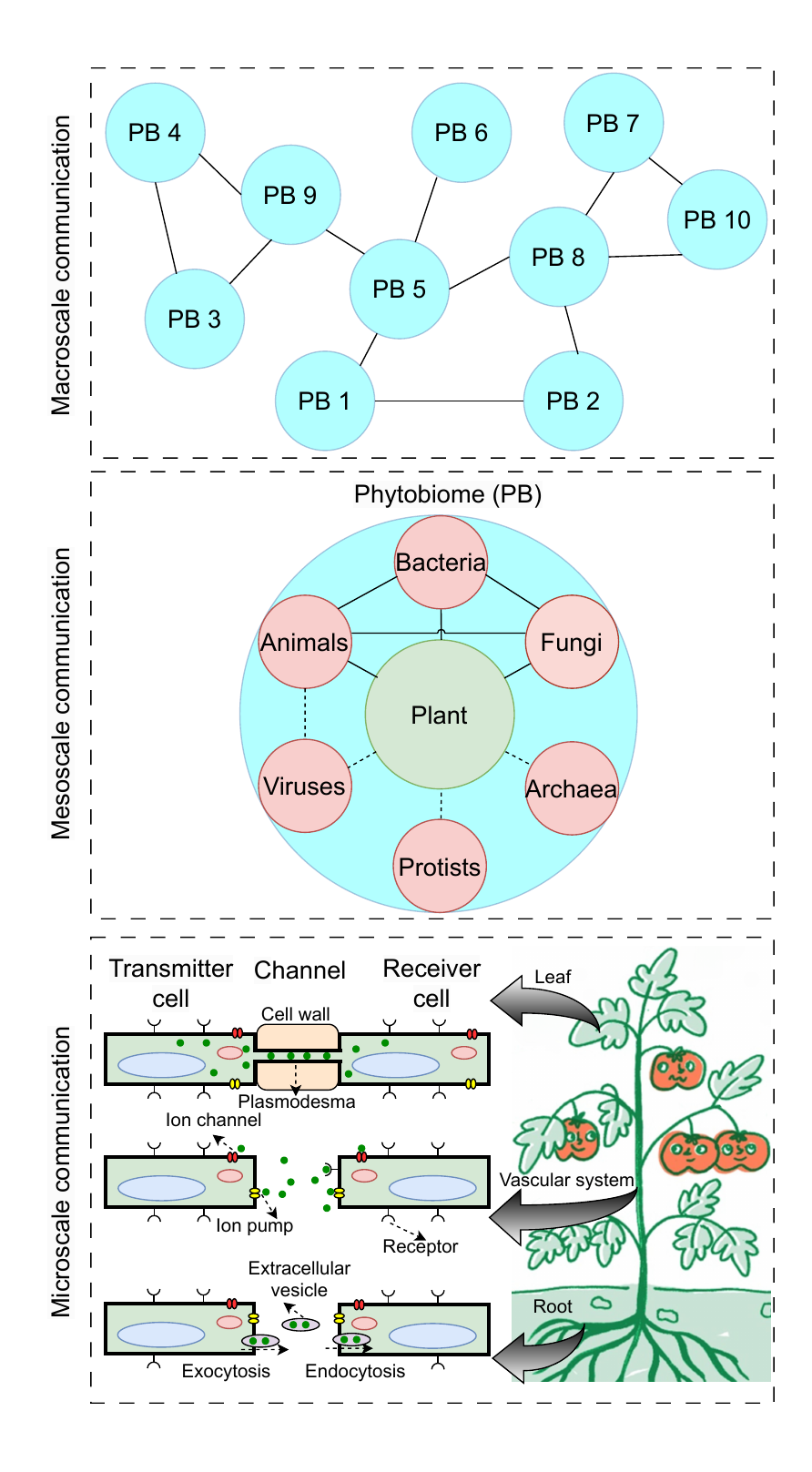}
	\caption{Multi-scale phytobiome communication framework.}
	\label{Network}
 \vspace{-0.7cm}
\end{figure}

It is well understood that organisms use information and communication to increase their evolutionary fitness. Moreover, the fitness value of information can be quantified with information-theoretic analysis, such as Kelly betting, which assesses how information contributes to survival and adaptation. For instance, the maximum growth rate of an organism such as a plant in an uncertain environment depends on the entropy of the environmental state \cite{donaldson2010fitness}. Communication in phytobiomes occurs across multiple scales, each with distinct dynamics. At the microscale, intra-organismal communication involves molecules like plant hormones transmitted through phloem tissues in response to stress. At the mesoscale, interkingdom interactions occur between plants and microbial communities via mechanisms such as bacterial QS and fungal communication at the root–soil interface. At the macroscale, communication spans across phytobiomes, e.g., stressed plants emit VOCs alerting neighboring plants. Analyzing phytobiomes as communication networks across these scales enables a deeper understanding of plant behavior and opens opportunities to design targeted interventions for enhanced agricultural efficiency, as detailed in Section~\ref{EPC}. This multi-scale perspective not only enables tractable decomposition of complex interactions but also aligns with the nested, modular organization of biological systems. To formalize this analysis, we propose a multi-scale communication framework illustrated in Fig.~\ref{Network}.


\subsubsection{Microscale Communication}
Microscale communication focuses on the direct information transfer between cells within the same phytobiome such as plant or bacterial cells. These cells can be both transmitter and receiver (or transceiver) similar to the basic components of the physical layer of conventional networks. Communication is achieved through molecule exchange between the transmitter and receiver cells via various mechanisms in different parts of the plant, e.g., leaf, vascular system, and root, as shown in Fig. \ref{Network}. 

As depicted in Fig.~\ref{Network}, transmitter cells release molecules via passive diffusion or active transport mechanisms, such as ion pumps/channels, or exocytosis in extracellular vesicles. In phytobiomes, the communication channels differ from conventional MC models due to the unique biological structures of plants. For instance, plasmodesmata directly connect adjacent plant cells, enabling targeted intercellular transport, while vascular tissues like xylem and phloem support long-range advection-based communication. Furthermore, receiver cells detect these molecules via plasmodesmata, receptor binding, and endocytosis. While the core dynamics can be modeled with the advection-diffusion-reaction framework \cite{jamali2019channel}, several plant-specific challenges must be considered. 



\subsubsection{Mesoscale Communication}
We define mesoscale communication as inter-kingdom communication among phytobiome members. As illustrated in Fig. \ref{Network}, the plant is the center of the phytobiome communication, similar to a hub in a local area network (LAN). While communication paths among the plant, bacteria, fungi, and animal kingdoms are well-documented (solid lines in Fig. \ref{Network}), paths involving archaea, protists, and viruses need further experimental studies (dashed lines in Fig. \ref{Network}). Such communication paths ultimately affect plant health. For example, our study in \cite{gulec2023stochasticd} shows that disrupting the MC between bacteria and plants can enhance plant health by disrupting the bacterial biofilm.  This MC perspective supports holistic system modeling of the phytobiome. Additionally, the analogy between communication networks and mesoscale phytobiome communication enables tools from network theory and science. For instance, connections in the phytobiome resemble a partial mesh network where the plant is the access point. Hence, tools from the graph theory can be used to reveal information propagation within the phytobiome.

\subsubsection{Macroscale Communication}
The networking among phytobiomes refers to macroscale communication at the top of the phytobiome communication framework as shown in Fig. \ref{Network}. Phytobiomes are connected through wired and wireless channels (Fig. \ref{Phytobiome}), analogous to wide area networks linking multiple LANs. Communication among phytobiomes can be analyzed and modeled using principles from communication network theory. For instance, optimizing network architecture to maximize throughput can lead to the design of more efficient agricultural fields. Furthermore, using graph theory, where phytobiomes are represented as vertices and their communication paths as edges, can help us understand how information flow influences ecosystem-level interactions, leading to more efficient use of water and agrochemicals. Next, the relation of  MC to electrophysiological signals is detailed.

\subsection{From Molecular to Electrophysiological Signals in Phytobiome Communication} \label{MEP}
\subsubsection{Relation of MC to Action Potential Signals}
Molecular exchange among plant cells generates electrophysiological signals through voltage changes across cell membranes. The flux of ions, such as Ca\(^{2+}\), through ion channels/pumps, as illustrated in Fig. \ref{Network}, causes depolarization and repolarization of the cell membrane. These voltage changes occur in slow (mm/s) and fast (cm/s) fashions, referred to as variation potential and action potential (AP) signals, respectively. We focus on APs due to their faster propagation speed. AP signals are induced by stimuli or stress. For example, in Venus flytrap, APs trigger rapid leaf closure to trap insects. Similar AP signals occur in fungi, indicating their importance for potential applications, such as fungal computers \cite{adamatzky2018towards}.

MC enables AP generation among cells. Hence, the MC perspective can be utilized to model these electrophysiological signals. For instance, the voxel model proposed by our group in \cite{awan2019communication}, where each cell is represented as a 3-D box and MC occurs through diffusion and active transport among cells, is employed to model APs. The application of communication and information theory allows us to quantify how reliably information is transmitted between cells via the mutual information between the transmitted signal and the receiver’s output signal. Additionally, the propagation speed is determined by measuring the time difference at which this mutual information surpasses a defined threshold. These insights help assess the efficiency of communication mechanisms in plants. Readers are referred to our work \cite{awan2019communication} for mathematical details.

\begin{figure}[t]
\begin{center}
\includegraphics[trim=0cm 0cm 0cm 0cm ,clip=true, width=0.7\columnwidth]{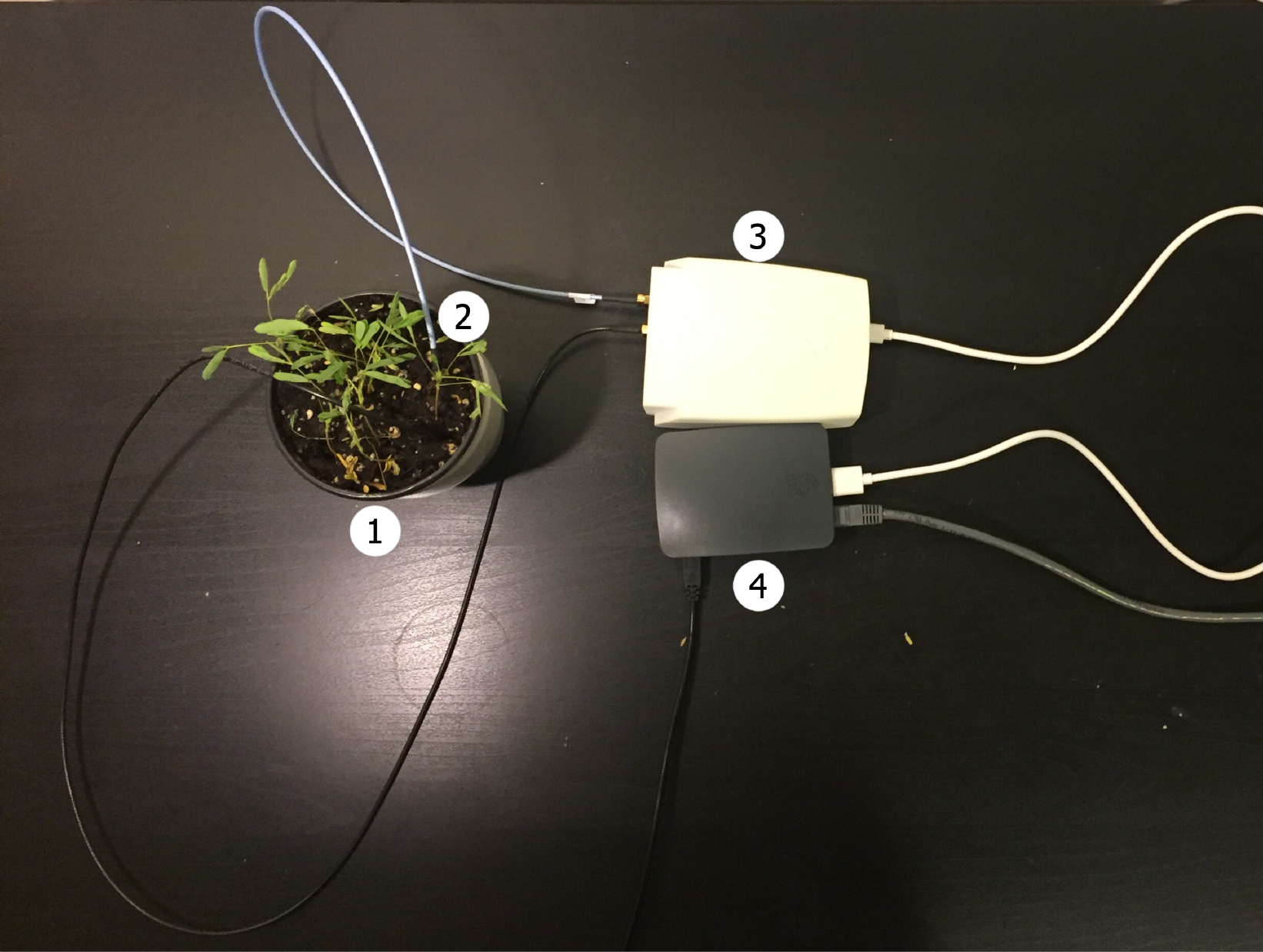}
\caption{Experimental setup for measuring AP signals in a \textit{Mimosa pudica} plant. 1) \textit{Mimosa pudica} plant, 2) Electrophysiological sensor, 3) Signal amplifier, 4) Single board computer. Adapted from \cite{awan2019communication}.}
\label{Experiment1}
\end{center}
 \vspace{-0.7cm}
\end{figure} 

\subsubsection{Experimental Validation}
To validate the MC-based voxel model of AP generation, we conducted experiments on \textit{Mimosa pudica}, also known as the sensitive plant closing its leaves when touched. Both the model and its experimental validation were introduced in our work \cite{awan2019communication}. As illustrated in Fig. \ref{Experiment1}, the experimental setup consists of an electrophysiological sensor connected to the plant's stem and soil, a signal amplifier, and a single-board computer to measure voltage changes in APs triggered by touch stimuli, which start the communication from the leaves to other plant organs, e.g., stem or root. Notably, the measurement of electrophysiological signals is prone to environmental interference, particularly from the electricity grid. To mitigate this source of noise, we applied notch filtering around \(50\) Hz.

\begin{figure}[b]
\begin{center}
\includegraphics[trim=0cm 0cm 0cm 0cm ,clip=true, width=0.8\columnwidth]{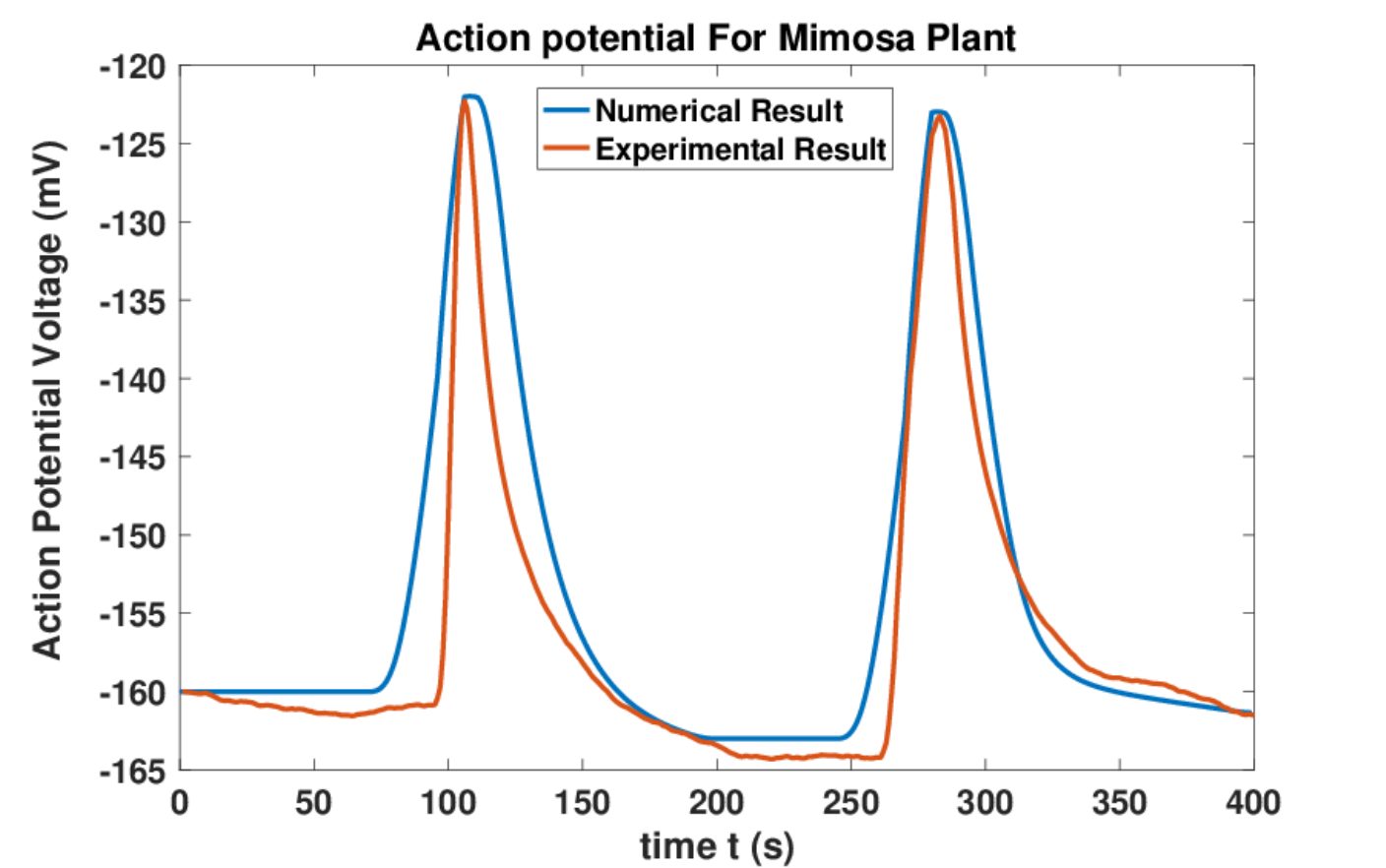}
\caption{Experimental and numerical results for AP signals, from \cite{awan2019communication}.}
\label{compare2}
\end{center}
 \vspace{-0.3cm}
\end{figure}

For multiple AP signals induced by two stimuli on the plant, the numerical results of the MC-based voxel model and measured signals, using the parameter values in \cite{awan2019communication}, are given in Fig. \ref{compare2}. These results of our proof-of-concept study validate the ability of MC to model electrophysiological signals and pave the way towards understanding the phytobiome communication. The next section elaborates on novel applications facilitated by engineering the phytobiome communication.


\section{Engineering the Phytobiome Communication} \label{EPC}
Adopting a communication engineering perspective not only allows us to understand the phytobiome communication but also provides opportunities to engineer it for agricultural applications.  In this section, novel applications to improve plant health and crop production efficiency are described.

\begin{figure*}[tb]
	\centering
	\includegraphics[width=0.7\textwidth]{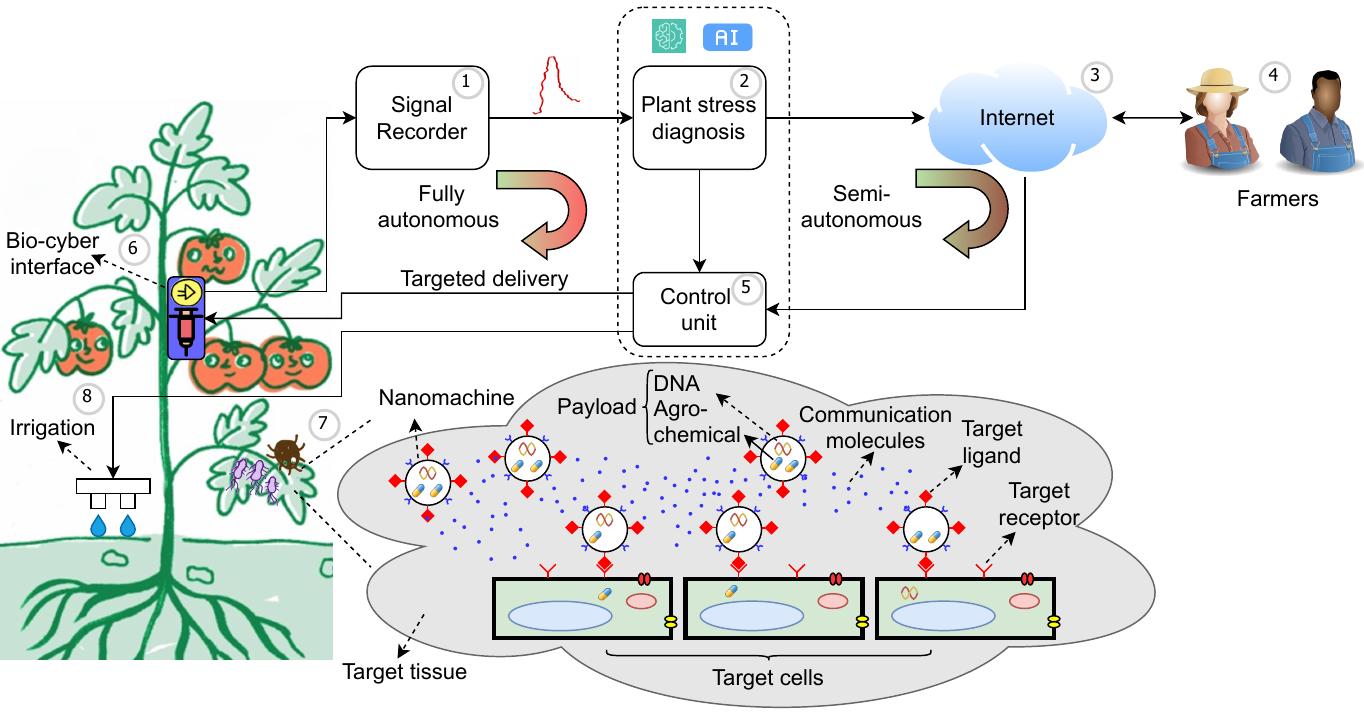}
	\caption{Applications: Phytobiome monitoring (steps 1-4), fully autonomous (steps 1, 2, 5-7) and semi-autonomous (steps 1-7) targeted delivery of agrochemicals/DNA via IoBNT, fully autonomous (steps 1, 2, 5, 8) and semi-autonomous (steps 1-5, 8) smart irrigation.}
	\label{Phytobiome_eng}
 \vspace{-0.3cm}
\end{figure*}

\subsection{Phytobiome Monitoring}
The first application leverages electrophysiological signals originating from intercellular communication within the phytobiome (see Section \ref{MEP}). Under stress conditions, e.g., drought, pest infestation, or nutrient deficiency, plants generate time-varying electrophysiological signals enabled by MC to coordinate responses such as the closure of stomata (tiny pores in the leaf's skin used for gas exchange) or defense activation. These stress-induced signals exhibit characteristic changes in amplitude, frequency, and temporal dynamics, which are already being analyzed by ML/AI techniques in both academic and industrial settings for early stress detection. Fig. \ref{Phytobiome_eng} (steps 1-4) illustrates this process, where electrophysiological signals are recorded via the sensor in the bio-cyber interface and then processed to diagnose the stress type. For example, in \cite{najdenovska2021identifying}, features extracted from recorded time-series signals were classified using the Extreme Gradient Boosting ML method, achieving an accuracy of $80\%$. Deep learning methods, such as multi-layer perceptrons and convolutional neural networks, were also applied in the literature. These diagnostic results can be shared with remote farmers through the Internet for smart agriculture applications. However, how mesoscale phytobiome communication alters these signals is still unclear and requires further research to improve classification accuracy using signal processing and ML methods.


\subsection{Targeted Delivery of Agrochemicals and Genes}
Here, we propose the usage of the IoBNT, which involves coordinated NMs connected to the internet, for the targeted delivery of agrochemicals and genes, as illustrated in Fig. \ref{Phytobiome_eng}. 

This application integrates phytobiome monitoring with delivery in two modes: fully autonomous and semi-autonomous. In the fully autonomous mode (steps 1, 2, 5, and 6 in Fig. \ref{Phytobiome_eng}), the injection of NMs is decided autonomously via ML/AI based on diagnostic results for plant stress types, e.g., nutrient deficiency, infection, or pest infestation. In the semi-autonomous mode (steps 1-7 in Fig. \ref{Phytobiome_eng}), the injection decision is made by farmers through an internet connection. Afterward, NMs with their payloads localize their target cells by using MC among themselves and bind to them via engineered ligands such as proteins and aptamers. These target cells can be plant cells hosting herbivores or pathogenic bacteria. NMs can be configured to track plant stress hormones like JA to locate affected regions. For instance, pesticides can be delivered to leaf cells where herbivores are located, or NMs can be designed to target receptors on the pests' surfaces for direct delivery. Fertilizers can also be similarly directed to nutrient-deficient cells through NMs. In both modes, the closed loop starting and ending in the bio-cyber interface enables continuous monitoring, allowing ML/AI methods to be fed with more data to increase the accuracy of stress diagnosis and injection decisions. Furthermore, gene transfer using modified DNA can enhance plant resilience. Nanoparticles can be employed to mediate DNA from NMs to target cells \cite{vega2020nanoscale}. 

\vspace{-0.3cm}

\subsection{Smart Irrigation}
In this application, a smart irrigation system, where plants autonomously manage water intake based on ML/AI-based stress diagnosis, is proposed. Similar to the targeted delivery application, there are two modes. In the fully autonomous mode (steps 1, 2, 5, and 8 in Fig. \ref{Phytobiome_eng}), the system autonomously activates upon detecting drought stress. Farmers can also control the irrigation via an Internet connection in the semi-autonomous mode (steps 1-5, and 8 in Fig. \ref{Phytobiome_eng}). This system continuously monitors the electrophysiological signals of the plant, providing a continuous training data flow to enhance the accuracy of ML/AI methods. The primary aim of this smart irrigation system is to prevent over- or underwatering, thereby increasing crop production and water use efficiency.

\subsection{Engineering Phytobiome Communication with Genetically Modified Bacteria}
Microbes (e.g., bacteria) in the human gut affect human health through the immune system. Similarly, microbes within the phytobiome influence plant health which can be improved by increasing the number of symbiotic microbes and enhancing their communication with the plant. Advances in biotechnology enable the use of genetically modified (GM) bacteria in MC applications. For instance, these GM bacteria can interfere with the MC of pathogenic bacteria and improve the MC between the plant roots and symbiotic bacteria, leading to improved nutrient uptake, disease resistance, and plant growth. However, GM applications should be pursued with consideration for biosafety and regulatory guidelines to ensure responsible deployment.

\subsection{Engineering Inter-Phytobiome Communication}
As shown in Fig. \ref{Phytobiome} and explained in Section \ref{DPC}, wireless and wired channels between phytobiomes are crucial to transfer information such as pest infestation. Plants receiving this information via VOCs or fungal networks can activate defense mechanisms to increase pest resilience. These VOCs, such as methyl jasmonate, can be synthetically produced and transmitted to neighboring plants via sprayer-based airborne MC systems for efficient information transfer. Communication parameters such as transmission time, modulation type, and signal power can be optimized by designing and analyzing this MC system. Additionally, in wired channels, there is potential to enhance communication by using synthetic fungi to improve inter-phytobiome communication. 

\section{Challenges and Open Research Issues}
This section highlights key open research issues and challenges for decoding and engineering phytobiome communication. As discussed in Section \ref{DPC}, a phytobiome is an intricate communication network with its unique challenges. At the microscale, these include the plant cell wall, which introduces an additional barrier; structural complexity of plasmodesmata; and time-varying flow in xylem and phloem. Notably, plasmodesmata exhibit dynamic gating in response to temperature, pressure, or pathogen attack, introducing time-varying noise and constraints on molecule propagation. Hence, tailored communication models are required to capture these distinct features and better understand how intercellular communication influences plant responses. 

At the mesoscale, experimental studies are needed to reveal the inter-kingdom communications among phytobiome members. This requires collaboration with plant scientists to decode this communication.  Moreover,  linking such inter-kingdom communication to electrophysiological responses will be essential for integrating them into the proposed ML/AI-based diagnostic framework shown in Fig. \ref{Phytobiome_eng}.

As detailed in Section \ref{EPC}, targeted delivery of agrochemicals in the phytobiome is a promising research direction, with ongoing MC research using magnetic nanoparticles offering a potential solution \cite{wicke2021experimental}. In addition, phytobiome-specific methods are needed to localize target cells with NMs. While the research on IoBNT is currently at an experimental stage, its adaptation to agricultural settings offers a lower-cost, more accessible alternative to biomedical IoBNT scenarios. On the other hand, plant stress monitoring via electrophysiological signals and ML/AI is at a more advanced stage in both academia and industry. However, it is imperative to investigate how the intricate structure of phytobiome communication in field conditions affects the accuracy of stress detection.

\section{Industrial and Agricultural Outlook}
For the last 30 years, increased agricultural efficiency has been driven partly by improved use of data in a collection of technologies often termed precision agriculture (PA)\cite{sishodia2020applications}. In PA, farmers collate information from the environment, satellites, drones, weather stations, soil moisture sensors, and cameras mounted on tractors. These data are then used to respond in a more precise way to local conditions. However, the future of agriculture lies in smart agriculture, which goes beyond PA by incorporating advanced technologies such as the IoBNT, ML/AI, and robotics to create a more interconnected, automated, and responsive farming system.

In smart agriculture, there is increasing awareness that the next step in efficiency will include gathering information from the plants themselves. This includes the use of technologies like chlorophyll fluorescence, hyper-spectral, and stomatal cameras to improve the temporal resolution of agricultural treatments. In modern high-tech greenhouses, climate computers adjust the conditions using a combination of mechanisms such as automated windows, heaters, and lighting.  Outdoor farmers also seek to understand how they can adjust their practices in a timely way. If farmers can receive reliable data from their crops via smart agriculture technologies, they can significantly improve yields while reducing environmental impact.  Examples include providing optimum levels of inputs such as water and fertilizers, as well as replacing pesticides with eco-friendly crop treatments.

At the forefront of this movement to help farmers are electrophysiological signals used for communication within the phytobiome. Understanding the communication mechanisms in plants with the use of ML/AI can give dramatically earlier insights into crop health. This can range from a matter of hours in the case of irrigation to many months for certain critical blights. This technology is particularly advantageous for root-borne diseases, which cannot be examined using optical techniques. Furthermore, it is notable that most crops appear to have similar electrophysiological characteristics, meaning that the interpretation of signals in one crop can have useful diagnostic applications in others. Additionally, the integration of IoBNT for targeted delivery of agrochemicals and genes, and engineering the intra- and inter-phytobiome communication into smart agriculture can revolutionize the agricultural sector with more sustainable and efficient practices.



\section{Conclusion}
In this paper, we explore the complex phytobiome communication and its use in smart agriculture. We propose a multi-scale communication framework to decode the interactions between plants and associated organisms in their environment through molecular and electrophysiological communication signals. This framework facilitates the understanding of the phytobiome as a communication network, and we review our experimental results to illustrate how it can model electrophysiological signals via MC. We then leverage this framework to engineer phytobiome communication with novel smart agriculture applications. Notably, the fully autonomous integration of smart irrigation and targeted agrochemical delivery via IoBNT, coupled with ML/AI-based diagnosis of plant stress from electrophysiological signals, holds the potential to eliminate human errors in sustainable, eco-friendly, and efficient agriculture. A practical near-term direction is to apply our framework to increase the accuracy of electrophysiology-based stress monitoring by investigating how phytobiome communication under field conditions affects signal patterns.


\section*{Acknowledgements}
The authors would like to thank Magdalena Janicka for the artwork in Fig. \ref{Phytobiome}. Fatih Gulec was supported by EU Horizon Europe under the Marie Skłodowska-Curie COFUND grant No 101081327 YUFE4Postdocs. The work of Andrew W. Eckford was supported by a Discovery grant from NSERC.

\bibliographystyle{IEEEtran}
\bibliography{IEEEabrv,ref_fg_phy}

\section*{Biographies} 
\vspace{-1cm}
\begin{IEEEbiographynophoto}{Fatih Gulec} [M]  received his Ph.D. degree in Electronics and Communication Engineering from Izmir Institute of Technology, Turkey, in 2021. He is currently an MSCA YUFE Postdoctoral Fellow with the Department of Computer Science and Electronic Engineering at the University of Essex, UK. His research interests include molecular communications, mathematical biology, and plant electrophysiology/bioacoustics. He was awarded a short-term DAAD Postdoctoral Fellowship in 2021, the MSCA YUFE Fellowship in 2024, and the 2022 Doctoral Thesis Award from the IEEE Turkey Section.
\end{IEEEbiographynophoto}

\vspace{-0.9cm}

\begin{IEEEbiographynophoto}{Hamdan Awan} [M] received his Ph.D. degree in computer science and engineering from the University of New South Wales, Australia, in August 2017. He is currently a lecturer of Computer Science, at Munster Technological University, Ireland. His research interests include molecular communications, information theory aspects of biological communication, and computer vision.
\end{IEEEbiographynophoto}

\vspace{-0.9cm}

\begin{IEEEbiographynophoto}{Nigel Wallbridge} received his Ph.D. degree
in medical engineering from the University of Leeds, UK, and his MBA degree from INSEAD, France. He is the executive chairman at Vivent, SA, Switzerland. 
\end{IEEEbiographynophoto}

\vspace{-0.9cm}

\begin{IEEEbiographynophoto}{Andrew W. Eckford} [SM] received his Ph.D. degree in electrical engineering from the University of Toronto in 2004. He is an Associate Professor with the Department of Electrical Engineering and Computer Science, York University, Canada. His research interests include the application of information theory to biology and the design of communication systems using molecular/biological techniques. His research has been covered in media, including The Economist, Wall Street Journal, and IEEE Spectrum. He received the 2015 IET Communications Innovation Award, and was a Finalist for the 2014 Bell Labs Prize.
\end{IEEEbiographynophoto}
\vfill

\end{document}